\documentclass[10pt]{article}
\usepackage{graphics,psfig}

\newcommand \ltw{\>\hbox{\lower.25em\hbox{$\buildrel <\over\sim$}}\>}
\newcommand \gtw{\>\hbox{\lower.25em\hbox{$\buildrel >\over\sim$}}\>}

\def\dcaption#1#2{\parindent=0pt%
     \footnotesize{\bf{Figure~#1}\quad\rm#2}}

\begin{document}

\setlength{\textheight}{9.0in}
\setlength{\textwidth}{7.0in}
\setlength{\oddsidemargin}{-0.3in}
\setlength{\evensidemargin}{-0.3in}
\setlength{\topmargin}{-0.2in}

\twocolumn

\title{Plasma physics in clusters of galaxies}

\author{Jean A. Eilek, Physics Department, New Mexico Tech, Socorro NM
87801}

\date{to appear in {\it Physics of Plasmas}, May 2003}

\maketitle

\begin{abstract}

Clusters of galaxies are the largest self-gravitating structures in the 
universe.  Each cluster is
filled with a large-scale plasma atmosphere, in which primordial matter is 
mixed with matter that has been processed inside stars.  This
is a wonderful plasma physics laboratory.  Our diagnostics are the
data we obtain from X-ray and radio telescopes. The thermal plasma is
a strong X-ray source; from this we determine its density and
temperature.  Radio data reveal a relativistic component in 
the plasma, and first measurements of the intracluster magnetic field 
have now been made. Energization of the particles and the field must be 
related to the cosmological evolution of the cluster.  
The situation is made even richer by the few galaxies in 
each cluster which host radio jets.  In these galaxies, electrodynamics 
near a massive black hole in the core of the galaxy lead to a collimated 
plasma beam which propagates from the nucleus out to supergalactic scales.
  These jets interact with
the cluster plasma to form the structures known as radio galaxies.  The 
interaction  disturbs and energizes the cluster
plasma.  This complicates the story  but also helps us understand 
both the radio jets and the cluster plasma.


\end{abstract}


\section{Overview}

Clusters of galaxies that we see today have been built up by gravitational
collapse of tiny density fluctuations in the early universe.  
The galaxies in the cluster swim in a sea of hot plasma.  Because
this plasma is easy to observe, and has been thought to be simple
to interpret, it seems an attractive probe of the cluster structure, and
thus cosmology.  However, we need to understand the plasma before we
can use it to study larger questions.

In this paper I review our current knowledge of the cluster
plasma, with emphasis
on areas of uncertainty and areas where input from other arenas of
plasma physics will be particularly helpful.  I will use
astronomical length scales. One parsec (pc) is about 3 light-years.
The diameter of a typical galaxy $\sim 30$ kpc.  The length of a radio jet
ranges from a few to a few hundred kpc. The diameter
of a cluster core  $\sim 0.5$Mpc, while the entire cluster extends out
to $\gtw 2$ Mpc.   An astronomical unit (AU) is the distance
from the earth to the sun, $\sim 5 \mu$pc.    The gravitational
radius of a massive black hole $\sim 1-10$AU.

\subsection{Clusters of galaxies}

Clusters of galaxies are the density maxima in the large-scale distribution
of mass in the universe.
The largest, brightest clusters are called ``rich'';
they are the easiest to study.  Much of the gravitating matter in any
cluster is ``dark'', meaning it does not radiate in any detectable
band;  it is thought to be non-baryonic. Some of the
luminous, baryonic matter formed the stars within galaxies; the rest
 remained as the diffuse plasma atmosphere of the cluster.  This Mpc-scale
plasma is easily detected with X-ray telescopes, and is 
commonly referred to as the Intra-Cluster Medium (ICM).   

\subsection{Jets from galactic cores}

A few galaxies within the cluster host Active Galactic Nuclei (AGN)
in their cores. The combined action of gravity and angular momentum
forces galactic matter into an accretion disk around
a massive central black hole.  Electrodynamic forces drive
jets of relativistic plasma out from the disk.  These jets, which begin
on $\mu$pc scales, remain collimated while propagating to much larger
scales (kpc to Mpc).   They are 
most often detected by their radio synchrotron emission, thus are
called ``radio jets''.  They usually escape their parent 
galaxy and interact with the local ICM. The beautiful structures formed
when the jets interact with the ICM are called ``radio galaxies''.   

\subsection{The plasmas in this environment}

In many ways the study of the cluster-wide plasma is still in its
infancy. We are gathering data and
trying to understand the basic physical state of the plasma. 
Much of the cluster plasma is well thermalized, obeying a
Maxwellian distribution function (DF) at 
warm, but subrelativistic, temperatures.
We know the plasma is magnetized, but as yet we know little about the
strength or structure of the field. The plasma may also contain
 a relativistic component, with a non-thermal DF,  in which the internal 
energy per particle is much greater than the rest mass.

The study of radio jets is somewhat more mature.  We know the
jet plasma contains highly relativistic particles, again with a
non-thermal DF. It is not clear
whether it contains any cooler or subrelativistic component.
The jet plasma is magnetized, probably at approximate equipartition
levels.  Current work in this field is aimed at understanding the
energetics and stability of these well-collimated plasma flows.

\section{Diagnostic tools}

This relatively new field  is driven by data taken with 
radio and X-ray telescopes.  These observations can be grouped into 
four types of diagnostics.  All apply to the cluster plasma;  only
one or two apply to radio jets. 

\subsection{Thermal X-ray emission}

This is our main diagnostic for the cluster plasma.  Several generations
of X-ray satellites
 have enabled detailed study of the thermal component of the plasma.$^{1}$

 Thermal bremsstrahlung from a plasma at density $n$ and 
temperature $T$ has emissivity, 
\begin{equation}
\epsilon_{brem}(\nu) \propto n^2 g(\nu,T) e^{ - h \nu / k T}
\end{equation}
where $g$ is slow function of frequency and temperature.$^{2}$

Broad-band X-ray images  of nearby clusters are well resolved spatially, 
showing that most rich
clusters are regularly shaped and quasi-spherical.  Figure 1 shows an
example. Straightforward deprojection
of the data tells us the density structure of the thermal ICM.
X-ray spectra can also
be obtained, with moderate spatial resolution, revealing spectral lines
from highly ionized heavy elements in addition to pure bremsstrahlung.
From these data we determine the temperature and composition of
the thermal ICM. 

\begin{figure}[htb]
\vspace{0.1in}
\centerline{ {\psfig{figure=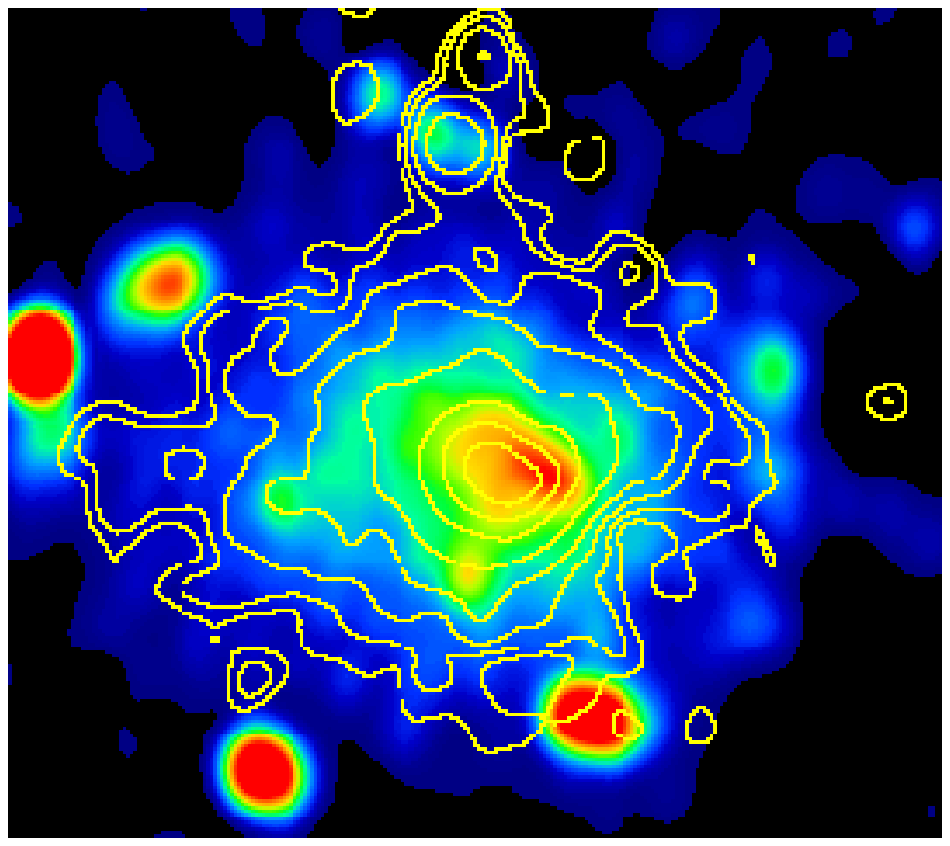,width=0.95\columnwidth}}}
\vspace{0.1in}
\dcaption{1} The regular cluster A2163.  Contours are from the X-ray
image, showing the distribution of bremsstrahlung from the
 hot, thermal plasma.  False-color image$^{8}$ is the radio emission, showing
diffuse synchrotron from the relativistic, magnetized component of
the plasma. The bright radio spots are individual cluster galaxies or
background sources.   The scale of the image $\sim 1$ Mpc.  
X-ray data from the ROSAT archive.
\end{figure}

\subsection{Synchrotron emission}

Thus is our main diagnostic  for radio jets, which are bright
enough that high quality images can be obtained with radio
 interferometers.$^{3}$  Synchrotron emission from some jets can also be
detected in the optical (and possibly X-ray) bands. 

In addition, synchrotron emission
has now been detected from the diffuse plasma in some clusters of
 galaxies.$^{4}$
When this emission extends throughout the thermal ICM, and is called a 
``radio halo''.  In other clusters, 
it sits on the edge of the cluster, and is called a  ``radio relic'';  
in still other clusters, the emission seems to be
patchy and irregular relative to ICM.

What can we learn about the emitting plasma? Synchrotron
emission comes from relativistic particles in a magnetized
plasma. The emission from a single particle peaks at 
$\nu \propto \gamma^2 B$. 
A useful way to write the emissivity is
\begin{equation}
\epsilon_{sy}(\nu) \propto B^{1/2} \nu^{1/2} n[\gamma(\nu;B)]
\end{equation}
if $n[\gamma(\nu;B)]$ is the number of particles at  energy $\gamma$ 
which radiate at $\nu$ in a magnetic field $B$.  
Thus,  detection of diffuse synchrotron emission is a direct detection of
magnetic field and relativistic electrons.

Two details will be useful below.  If we assume the usual power-law
electron DF, 
$n(\gamma) \propto \gamma^{-s}$, we find a power-law radiation spectrum: 
$\epsilon_{sy}(\nu) \propto  \nu^{-(s-1)/2}$.  If we further
assume $s \simeq 3$, which is typical of cluster radio haloes, we
find $\epsilon_{sy}(\nu) \propto p_{rel} p_B \nu^{-1}$.  
The emissivity in this case depends on
the product of the pressures in relativistic particles and field. 

\subsection{Faraday rotation}

Polarized synchrotron emission from radio galaxies in clusters
allows us to measure the magnetic field  in the plasma between 
us and the galaxy.

The different phase speeds of right-hand and left-hand polarized 
signals, in a magnetized plasma, result in rotation of the 
plane of linear polarization as the signal propagates through the
plasma.   The polarization angle $\chi$ rotates by
\begin{equation}
{ d \chi \over d \lambda^2} \propto \int n \mathbf{B} \cdot d
\mathbf{l}
\end{equation}
Because of the algebraic sign in the inner product, and the possibility
of an $(n,B)$ anticorrelation, the integral in (3) can be much smaller
than the simple product $n |B| l$.  Thus, 
detection of Faraday rotation is a detection of
magnetic field,  but the converse does not hold.

If the plasma responsible
for the Faraday effect is separate from  the emission source,
the signal will obey $\chi \propto \lambda^2$;  if the radio-loud plasma
itself causes the rotation, the simple $\lambda^2$ behavior will be
lost.  Faraday measurements of radio galaxies within or behind clusters
of galaxies usually show the clear $\lambda^2$ behavior, supporting
their use as probes of the ICM. 

Radio sources  in clusters generally have larger Faraday rotation 
than those not in clusters.  Imaging studies of embedded sources find
the rotation is ordered, not random, with order scale $\sim 3 - 30$ kpc.
High rotation is  detected from the dense, central ``cooling cores'' (see
below), while
lower rotations  are found from sources in lower density
 environments.$^{5}$

\subsection{Nonthermal X-ray emission}

X-ray spectra of the ICM show a hard X-ray (HXR) ``tail'' on the thermal
bremsstrahlung spectrum.   This
has only been confirmed for a few clusters,$^{6}$ but detecting
such a faint signal is difficult even with the new X-ray instruments
available. In addition, extreme-ultraviolet (EUV) emission has been 
reported from several clusters, but these are also difficult observations,
and the detections remain controversial.$^{7}$
The paucity of captured HXR or EUV photons has not, however, inhibited
modeling and speculation on the nature of this emission.  

The most attractive explanation is that this high-energy 
 emission arises from
inverse Compton scattering (ICS) of microwave background photons 
(relic radiation from the very early universe, now cooled to 2.7K) by
relativistic electrons. The emission from a single particle peaks at 
$\nu \propto \gamma^2 \nu_o$, if $\nu_o$ is the frequency of the incoming
photons.   The ICS emissivity per particle, in a radiation field with
energy density $U_{rad}$, can be written,
\begin{equation}
\epsilon_{ics} \propto U_{rad} ( \nu_o \nu)^{1/2} n[\gamma(\nu; \nu_o)]
\end{equation}
As with synchrotron, if we assume $n(\gamma) \propto \gamma^{-s}$, we
again get a scattered power law:  $\epsilon_{ics}(\nu) \propto \nu^{-(s-1)/2}$.
Because $U_{rad}$ is fixed, detection of ICS  from a cluster
is a direct detection of relativistic electrons in the cluster plasma.

\section{The intracluster plasma}

From the X-ray luminosity and spectra, we know that
the ICM is dominated by a hot, thin maxwellian plasma, at $n \sim 10^{-4} -
10^{-3}$cm$^{-3}$ and $T \sim 10^7 - 10^8$K.  It is almost all hydrogen,
with heavier trace elements at approximately half solar abundance.
The temperature and composition are approximately uniform throughout
the cluster, with only small spatial variations. The density is centrally
enhanced, with a central core radius $\sim 0.1-0.3$ Mpc, and falls
off at larger radii, as $\propto r^{-2}$ or faster.  The cluster A2163
in Figure 1 is a good example.  

From the composition of the ICM we infer its origins.  The very
lightest elements (H, He, and very small amounts of Li and Be) 
are primordial --
relics of nuclear reactions in the very early universe -- while the
heavier elements are 
created inside stars.  It follows that about half of the ICM
is primordial, having taken part in the original gravitational collapse of 
the dark matter.  The other half 
of the ICM must have been processed by stars in the cluster galaxies.
Massive stars are short-lived, and at death recycle most of their mass to
the interstellar medium in their galaxy.  Much of this material must
have been stripped or ejected from the galaxies, and joined the diffuse
ICM, in order to account for the current chemical composition of the
cluster plasma.

\subsection{Dynamical state}

It was initially thought that clusters of galaxies are static, 
self-gravitating systems, decoupled from the larger scale universe.
We now know this is only approximately true;  but it is still useful
as a ``zero-order'' description.  

Most rich clusters have smooth, centrally
concentrated distributions of gas and galaxies within the cluster core. 
The thermodynamics of the ICM seem
simple.  It was heated in the initial gravitational collapse of
the cluster.  Lacking present-day heating sources, the
ICM remains at that temperature today because radiative
losses are unimportant over the life of the cluster.
The dynamical state of the ICM also seems very simple. The dark
matter determines the gravitational potential.  The ICM sits in
 hydrostatic equilibrium in the potential well of the dark
matter:
\begin{equation}
\nabla^2 \Phi =  4 \pi \rho_{dark}~; \quad 
\nabla p_{icm}  = -  \rho_{icm} \nabla \Phi
\end{equation}
Simple solutions to the first equation show a characteristic core radius
of the gravitating matter, which depends on its central density and
internal energy.  Simple solutions to the second equation find
the core radius of the dark matter is also the core radius of the
cluster plasma. 

\subsubsection{Cooling cores}

An interesting subset of rich clusters deviate from this simple picture;
they have an unusually dense central concentration of plasma.  Radiative
cooling is important in these cores.  Unless there is
 ongoing heating, the central gas should be quite cool.  In addition,
the simple system (5) no longer describes the central ICM.  Instead
one expects radiatively regulated inflow,  with the gas slowly
settling into the center as its pressure support is lost.   Simple
models predict typical inflows of several hundred solar masses per
year in  strong cooling cores.$^{9}$

This overly simple picture has not been supported by observations.$^{10}$
Extensive searches  have failed to find the large amount of cooled
matter 
predicted to have accumulated in the cores (either as cold plasma
or as recently formed stars).   In addition, we now
know that radio-loud AGN exist in almost every cooling
core;  their jets may be important in heating the central ICM and
offsetting the radiative losses.

\subsubsection{Mergers:  flows,  shocks and cool fronts}

We now know that clusters are still evolving, as
small and large clumps of matter continue to merge with the system.
We believe minor mergers are common in most clusters at the present time.
Many clusters show subtle signs of this, such as bimodal galaxy
distributions, or small irregularities in the plasma distribution.
  Simulations$^{11}$ show that large-scale
bulk plasma flows and small-scale turbulence should be common;
unfortunately these are hard to measure directly.

A few clusters are undergoing major mergers, in which two mass clumps
of comparable size are merging at the present day.  These have clear
observational signatures.  Perhaps the most striking are the large,
peripheral shocks created when the two clumps hit supersonically. 
A2256, shown in Figure 2, is a striking example.
  The shock can be seen in the
X-ray image as a bright, transverse feature on the edge of the cluster.
Temperature maps show this feature is hotter than the rest of the cluster
plasma.$^{12}$
  Radio maps show this feature is bright in synchrotron
emission (see below), and that its magnetic field is ordered, parallel to the
bright emission ridge$^{13,14}$.  Both results
support the identification of this structure as
a very large ($\sim 0.5 $Mpc) plasma shock.

\begin{figure}[htb]
\vspace{0.1in}
\centerline{ {\psfig{figure=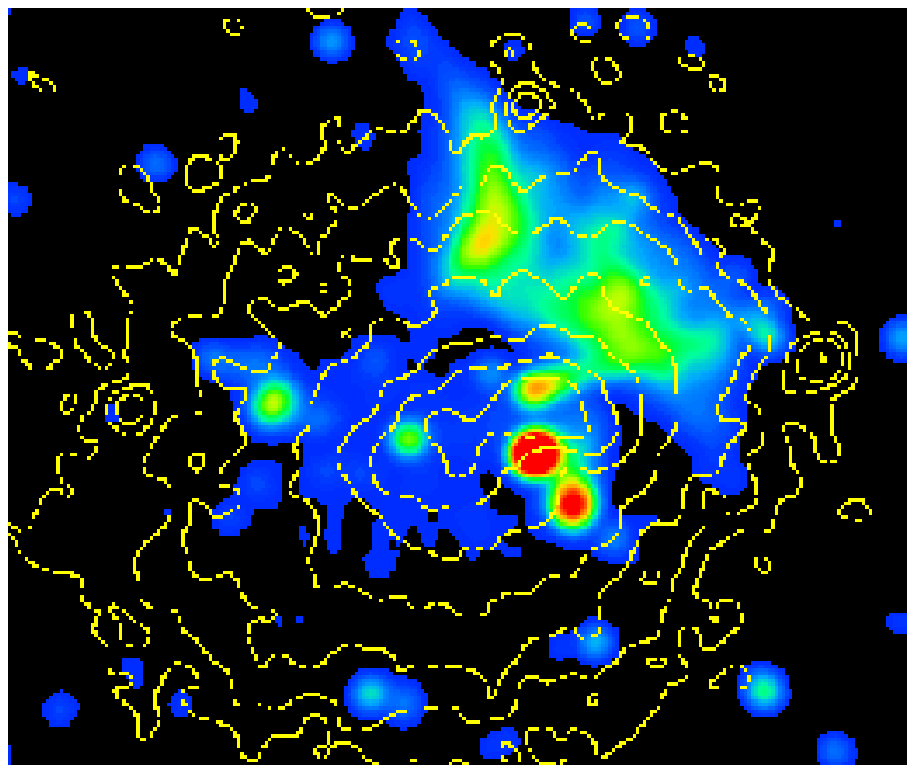,width=0.95\columnwidth}}}
\vspace{0.1in}
\dcaption{2} A2256 is an example of a merging cluster. False-color 
shows radio synchrotron emission from the nonthermal component.$^{14}$
 Contours show X-ray emission from the thermal plasma.
The thermal plasma shows some
distortion, to the upper right, where the plasma is hotter.  The radio
emission, however, reveals a large, offset feature which is thought
to be a shock created by the merger.  A fainter radio halo, central to
the cluster, can also be seen, as well as individual radio galaxies within
the cluster core.   The scale of the image $\sim 1$ Mpc.
X-ray data from the ROSAT archive.
\end{figure}

Recent X-ray data reveal new,  quite unexpected features, now called
``cool fronts''.$^{15}$  Some
merging clusters contain very thin, large-scale contact discontinuities, 
unresolved at a few kpc,  
between cool and warm plasma.  Their concave shape initially suggested a
bow shock, but the sign of temperature jump was found to be 
wrong.  Plasma behind the front turns out to be
cooler, not hotter, than that ahead of the front.  These are now
believed to be due to subsonic mergers, in which the cooler core of one
cluster is moving through warmer gas in another cluster, rather like
slow fluid  flow around a smooth rock.

\subsection{Magnetic field}

We are learing that the ICM is magnetized, as we detect 
both synchrotron  emission and  Faraday rotation from the plasma.  
Unfortunately, neither diagnostic allows clear 
determination of the strength or structure of the cluster field.

\subsubsection{Derived from radio haloes}

Diffuse synchrotron emission traces both the magnetic field and relativistic
particles in the ICM.  The intensity of the radiation depends on both
components;  without additional information we cannot determine the 
energy density in the field or the particles. One common approach in radio
astronomy is to minimize the total nonthermal pressure ($p_{rel}+p_B$),
subject to the constraint of the observed emissivity (as in equation 2).
Because this analysis leads to comparable field and particle energies,
it  is called ``equipartition''.  If HXR emission is also detected from
the cluster, its strength can be used to find the relativistic electron
density (from equation 4), allowing the field to be determined from the
synchrotron emission.  When a uniform, homogeneous magnetic field
is assumed, both methods tend to find $B \sim 0.1 - 0.5 \mu$G.$^{16}$

Radio halo observations provide only indirect information on the
field structure. Although synchrotron emission from a uniform
magnetic field is highly polarized, the haloes are unpolarized. 
It follows that several field
reversals must occur along the line of sight within the cluster, or
transversely within the resolution of the observations.  From 
this we can infer the field order scale is no larger than 
$\sim$ tens of kpc.

\subsubsection{Derived from Faraday rotation}

In order to use Faraday
 data to measure the cluster field, one must know
the structure of the field and the degree to which it mixes 
with the cluster plasma (equation 3). Most 
work has made the simplest assumption, that the cluster field is
disordered,
space-filling and well mixed with the thermal ICM. It is also common to
identify the order scale apparent in the RM images, $\sim 10$ kpc, 
with the order scale of the magnetic field.  This approach 
finds fields $\sim$ few $\mu$G throughout the cluster core, and
higher fields (tens of $\mu$G) in the dense cooling cores.$^{17,18}$ 

However, it is not clear if the magnetic field
inferred from RM data reflects the field throughout the cluster. 
There may be a boundary layer in which magnetic
field carried out by the radio jet mixes with the ICM. 
In addition, the radio 
jet will deposit energy to the local plasma, thereby driving turbulence
and amplifying smaller, seed fields there.   If either
of these effects is important, the rotation measures
 may not be typical of conditions throughout the cluster.
Statistical studies of radio sources  behind
(not inside of) clusters could in principle detect ``uncontaminated''
cluster field, but such studies have so far have been 
inconclusive.$^{18,19}$

\subsection{Relativistic plasma}

We are learning that the ICM can contain a significant fraction of highly
relativistic particles;  in some clusters we  have detected 
synchrotron and HXR emission from these particles.  We do not yet
know how common, or how significant, this component is in all clusters.

\subsubsection{A significant component?}

As yet we know very little in detail about this relativistic
plasma.  We do know that electrons at Lorentz factors $\sim 10^3
- 10^4$ are required to account for the synchrotron emission,
 and also to produce the HXR emission by Compton
scattering of the microwave background. The electron DF
 is not known with certainty.  By analogy to
galactic cosmic rays and to radio jets, the particle DF in the haloes
is often assumed to be a power law, {\it with an unknown low-energy 
cutoff}.  The energy density of the relativistic
electrons is at least a few percent of that in the thermal plasma, in
clusters with detected HXR. 
It could be larger if the low-energy cutoff is chosen
less conservatively.  We have no information on relativistic ions in
the cluster plasma.   Galactic cosmic rays contain
$\sim 100$ times more energy in baryons than in leptons.  If this
is also true for the cluster, the relativistic component of the
plasma is indeed important energetically. 

\subsubsection{A universal component?}

It is not yet clear if the ICM in every cluster has a 
strong relativistic component.
Because very few haloes or relics were initially detected,
they were thought to be rare.  If this is the case, they must be 
due to an unusual recent event in the cluster's life, such as 
a major merger.   On the other hand, more haloes are being
found, and 
the data available at present  hint at a correlation between
synchrotron and X-ray power in  rich clusters.$^{3,20}$  
If this trend is verified when more data become available, it 
strongly suggests that the nonthermal component is common to all
clusters, and probably simply a by-product of cluster ``weather''.
Our  uncertainty  reflects the fact that the observations
necessary to detect diffuse synchrotron emission from the cluster
plasma are difficult and time-consuming;  they are only beginning to
be carried out systematically.$^{21}$  We must wait a few years,
until work currently underway has been carried out, before we can say
whether diffuse synchrotron emission is common to all rich clusters,
or is found only in a special few.

\subsubsection{Origins and energization}

If radio haloes and relics turn out to be common, then 
the relativistic electrons have a lifetime problem. 
Their radiative lifetime (to ICS on the microwave background) 
is about one percent of the age of the cluster. 
Because radio haloes extend throughout the cluster volume, 
their electrons cannot have been injected or accelerated at some special
point in the cluster (such as an active galaxy or a large-scale merger
shock). Their diffusion rate is too slow, and for electrons is limited by
the radiative lifetime;  we would expect the radio emission to be localized
around the injection point.  Thus, relic sources may be due to one localized
event, but halo sources cannot be.  The electrons in radio haloes 
must be undergoing {\it in situ} energization  in the diffuse 
cluster plasma.

\subsection{An example:  the Coma cluster}

There are not yet enough well-measured halo clusters to draw broad
conclusions about the nonthermal plasma.  We can, however, look at
one nearby, well-studied object.  The thermal plasma in the
Coma cluster has a smooth, regular structure. Its core has
radius $\sim 300$ kpc, $n = .003$cm$^{-3}$, $T \sim 9 \times 10^7$K$^{22}$. 
It has detected HXR emission (whose spatial extent is unknown).
It has a smooth radio halo, extending throughout thermal gas halo. 
Radio galaxies within the core have Faraday rotation values typical of
other rich clusters, although sources in this cluster
have not been studied in as much detail as elsewhere.$^{17, 18, 23}$

The simplest model of the ICM in Coma would be a uniform, space-filling
magnetic field, fully mixed with the thermal and nonthermal plasmas.
With this assumption, one can determine $B$ from the ratio of ICS to
synchrotron power (equations 2 and 4);  the result is $B \sim 0.2 \mu$G.
This disagrees, however, with the Faraday data.  If the field has
an order scale $\sim 10$ kpc (consistent with other Faraday studies and
the lack of polarization), and is well mixed with the thermal plasma,
we need $B \sim 2 \mu$G.  This factor of ten discrepancy is unsatisfying
(even in astrophysics), and suggests the model is wrong.  In fact,
the uniform, space-filling magnetic field is physically
unlikely.  We know magnetic fields elsewhere (such as turbulent or 
 space plasmas) are inhomogeneous, with high-field, high-current
filaments, placed intermittently throughout a lower-field region$^{24}$. 

Consider, therefore,  a two-phase
plasma as our toy model. Let high-field flux ropes be surrounded by
low-field interfilament regions, with the two phases in pressure balance.
We want the filaments to have a covering factor at least unity (so that
any line of sight will show Faraday rotation).  The volume filling factor
must be $\sim 0.1 - 0.5$ for the 
length scales (10 \&  300 kpc) used in this toy model.  The filaments
can account for the Faraday signal if they are internally magnetized,
with  $B \sim 30 \mu$G and $p_{th} \ll p_B$ inside the filament.
Only a small fractional pressure in relativistic
electrons is required also to account for the radio halo.
The HXR is dominated by relativistic electrons in the interfilament
region.  The  minimum energy density needed there, 
to explain the HXR emission, is 
$\sim 1\%$ of the thermal plasma energy density. As discussed 
above, this value could be much larger.

\section{The plasma in radio jets}
 
Radio jets are well-collimated, relativistic outflows, which begin
on black-hole scales (several $\mu$pc), 
and propagate to extragalactic scales.
Bright features (waves or shocks) in the inner ($\ltw$ kpc) jet
show high-speed outward motion, at bulk Lorentz factor $\sim 3-10$,

We detect jets by their synchrotron radiation.  We therefore know
they contain a magnetized, internally relativistic plasma.  This radiation
is strongly polarized, from which we deduce the 
field is well-ordered.   This is our only diagnostic: there is
no internal Faraday rotation, nor any compelling
evidence as yet that the HXR seen from a few jets are due to ICS.
High resolution images show the plasma is inhomogeneous, 
with luminous filaments that probably trace high-$B$ regions.  
  There are hints 
that the plasma is mostly or fully relativistic, with no signficant
 cooler, thermal component.  Indirect arguments suggest the plasma 
is within  a factor $\sim 10$ of equipartition between the magnetic
field and relativistic electrons.   The equipartition fields are typically 
$\sim 10 - 100 \mu$G on kpc-Mpc scales, and larger in the galactic core.

Models of jet creation suggest even more extreme conditions might
exist.$^{25}$  Electrodynamic or MHD acceleration is the most likely close
to the black hole.  Such jets may carry a net current from the 
galactic nucleus out to extragalactic scales. The return path is not 
clear, but it is probably established through the interaction of the 
jet with the ambient plasma.  Some models start with a strong Poynting flux
jet; subsequent pair creation will produce an electron-positron jet.
Alternatively, mass-loading by entrainment from the ambient plasma
may turn these into electron-ion jets by the time we observe them on
extragalactic scales.

\subsection{Energetics}

Two issues deserve mention here.  How much energy does the jet
deposit in the ICM?  How does the jet convert flow energy
{\it in situ} to internal energy of the relativistic particles and
magnetic field?

\subsubsection{Energy carried by the jet}

What energy does the jet carry?  We can write it down easily enough:
\begin{equation}
P_j = \pi r^2 \gamma^2 \beta c \left( \rho c^2 + 4 p + { B^2 \over 4 \pi}
\right)
\end{equation}
assuming an internally relativistic plasma and using ideal MHD. But
 $P_j$ can only be determined indirectly in most cases. The luminosity
of the associated radio galaxy, which is easy to observe, is 
at best an indirect probe of $P_j$.  It depends on the
particle and field history of the source as well as current
conditions in the jet.
Nonetheless, estimates such as $P_{rad} \sim 10^{-2} P_j$ are common
in the literature.  Such estimates may be useful statistically, but they
can hardly describe a single source at any given time. 
Going beyond this, some authors have
invoked simple dynamical models of the larger radio source, for instance
what $P_j$ must have been in order for the jet to propagate a given
distance in a given time.  Such estimates are unsatisfying because 
they rely on important issues which are hard to determine,
such as source age or AGN duty cycle.

\subsubsection{Internal Energization}

Some of the energy flowing in the jet must be converted {\it in situ}
to relativistic particles and magnetic field.  This is needed to offset
both expansion and radiative losses.

The jets expand dramatically between their origin at the
accretion disk  and when they become observable as radio jets.
They continue to expand, albeit more slowly, out to much larger,
extragalactic scales.  
If the flow were adiabatic, the plasma would cool during
the expansion and the B field would  decay by flux freezing.  The observations
clearly contradict this:  the synchrotron emissivity does not decay
significantly going along the jet.$^{26}$

Radiative losses are also relevant. If the magnetic field is not too far
from  the equipartition value, the radiative  lifetime of the 
relativistic electrons is less than the travel time along the jet.
This  predicts a gradual steepening of the spectrum, going along the
jet (as the particles lose energy), followed by a frequency-dependent
 end to the emissivity. This again contradicts the observations.
Thus it seems very likely that {\it in situ} energization is taking
place in radio jets on $\gtw$ kpc scales.

\subsection{An example:  M87}

This galaxy, in the core of the Virgo cluster,
 is one of the closest AGN.  Synchrotron emission from its 
jet (shown in Figure 3)
has been detected in optical, radio and X-ray bands.$^{27,28}$
What do we know about this jet?

\begin{figure}[htb]
\vspace{0.1in}
\centerline{ {\psfig{figure=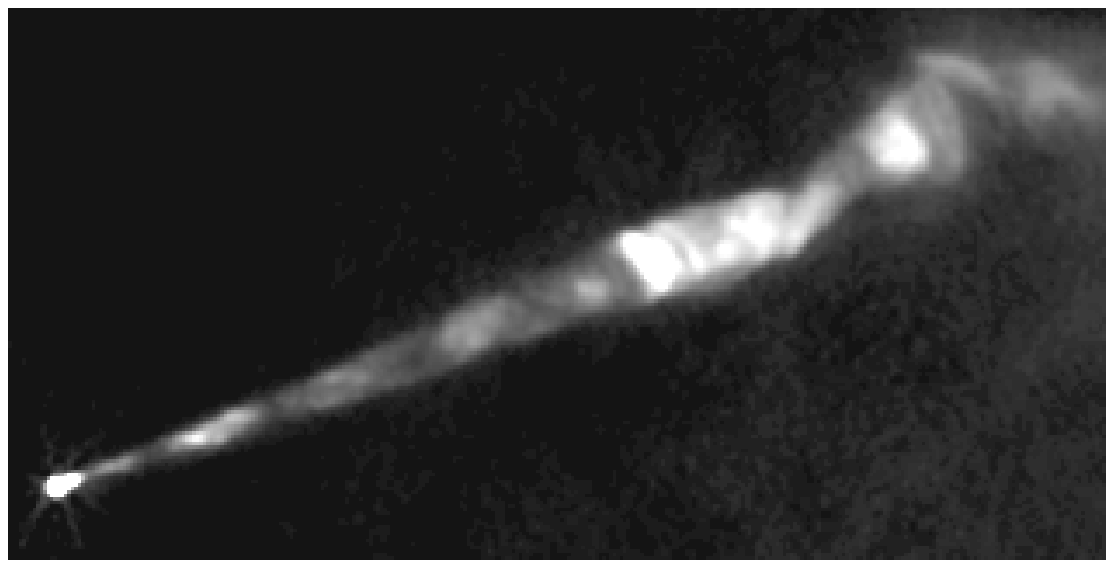,width=0.95\columnwidth}}}
\vspace{0.1in}
\dcaption{3} The jet in the nearby active galaxy M87, shown in a radio
image from the VLA$^{27}$. The bright spot at left is coincident with
the massive black hole in the galactic core.  
Most of the bright features show relativistic proper motions away 
from the core$^{29}$.  The scale of the image $\sim 1.5$ kpc. 
\end{figure}

Some answers come from direct measurement.  Well resolved radio and
optical images show significant internal structure, bright knots and 
filaments,  within a jet which
initially expands slowly, then collimates at a bright feature which may be
a transverse shock.  Images taken several years apart reveal
temporal variability of bright knots, as well as
 systematic outwards motion at $\gamma \sim 3$.

Some answers come from basic synchrotron theory.  The minimum-pressure
field $B \sim 250 \mu$G along the jet.  The electron DF extends from 
$\gamma \sim  2000$ (radio-loud particles) through $\gamma \sim 10^6$ 
(optical-loud), and probably up to $\gamma \sim 2 \times 10^7$ (X-ray loud).
If the underlying $B$ field is not too variable, the synchrotron spectrum
reveals the electron DF, which can be approximated as a broken power
law over this energy range. We can also derive a reliable
 lower bound to jet power from this analysis.  Referring to (6),
$P_j \gtw 4 \pi r^2 \gamma^2 \beta p_{min} c 
\gtw 3\times 10^{44}$erg/s, which is larger than the X-ray luminosity
from the inner core of the Virgo cluster.$^{30}$

Further analysis is possible in the context of a physical model. 
A ``double helix'' structure is apparent in both optical and radio
images of the jet.  It can be identified as a combination of helical
and elliptical wave modes.  If these modes are 
caused by the Kelvin-Helmholtz (KH) instability
and are close to internal resonance (the maximum growth rate), their
measured wavelengths and speeds allow determination of the plasma parameters.
With such analysis, one can estimate the jet Mach
numbers (both internal and external).  We$^{31}$ estimate the sound
speed in the jet plasma to be $c_s \sim 0.2-0.5 c$;  thus the
internal plasma is quite hot but not highly relativistic.  It must contain
a cooler component in addition to the very energetic particles which make
the synchrotron radiation.  The external plasma is cooler,
$c_s \sim .05 c$, and thus at higher density than the jet, 
but not as cool or  dense as the central ICM in Virgo.
This suggests that the plasma from the
radio jet has partially mixed with the ambient cluster plasma. 
New X-ray images$^{31}$ 
support this, revealing that
the local plasma is disturbed and disordered on this scale. 

\subsection{Jets in the ICM}

When the jet propagates to extragalactic scales, it interacts with
the ambient cluster plasma and forms the beautiful structure known
as a radio galaxy (RG).  We expect this interaction to heat the cluster
plasma; eventually the plasma and field carried by the jet
should be deposited in the cluster plasma.  Just how this will
happen depends on how the jet propagates into the ICM and whether
it is subject to disruptive instabilities.

\subsubsection{Propagation and stability}

Radio jets are  subject to several potentially disruptive
instabilities.  We know jets propagate at high  speeds into
the cluster plasma.  They will therefore be subject to shear-driven MHD
instabilities.   Extensive analysis of the KH instability
has been applied to jets, including magnetized jets and relativistic
jets.$^{33}$  This work suggests that weakly magnetized jets are
most stable when they are supersonic, but that poloidally magnetized
jets are most stable when they are subalfvenic. In addition, a jet which
is denser than its immediate surroundings tends to be more stable.  
One might speculate that local changes in the Alfven mach number, or
ambient density, trigger instability in a propagating jet. 

In addition, jets may carry a net current. Although this has not received
as much attention in the literature, some authors have suggested it$^{34}$,
as do most MHD or ED models of jet formation.$^{25}$  Current-carrying
jets are subject to another class of instabilities.  While such instabilities
have been extensively studied in laboratory plasmas, they have received
less attention in laboratory jets.  Some interesting
recent work$^{35}$ suggests that the kink instability will be the
most important for radio jets, but that instead of fully disrupting
the jet it may lead to increased internal dissipation and expansion
of the jet, followed by recollimation.  An interesting further speculation
is that the jet may undergo magnetic relaxation, perhaps triggered
by an instability, and evolve into  a minimum energy state when it reaches
large scales.$^{36}$

\subsubsection{Three types of sources}

Given the range of possible instabilites, one wonders how a jet
can manage to propagate at all.  This is
not yet solved; neither analytic theory or numerical simulation
 can address the entire issue.  Some
insight may be gained from a qualitative study of source
morphology.  Radio galaxies can be divided into three classes, based
on their morphology, which have some correlations with radio power and
parent galaxy size.$^{3,37}$ These three 
classes seem to reflect the development of instabilities in the jet.

\paragraph{Jets which remain stable}
Consider a jet which is not disrupted by any fluid or
current instability (for instance a highly supersonic fluid jet).
The rate at which its leading edge propagates into the ICM is
governed by momentum conservation.  Its advance speed may
 be supersonic with respect to the  ICM, but subsonic with respect 
to the jet plasma, causing  two shock
transitions.  A bow shock propagates into the ICM, and 
a quasi-transverse shock stands close to the end of the jet.  This
latter shock can be a site of strong, local particle acceleration,
as well as magnetic field amplification, thus producing the radio bright 
spot characteristic of many classical-double
RGs (Figure 4 shows one example).  The jet plasma which has passed 
through this shock, and decelerated, will 
 expand laterally, forming the ``lobes'' which characterize this type
of RG. 

\begin{figure}[htb]
\vspace{0.1in}
\centerline{ {\psfig{figure=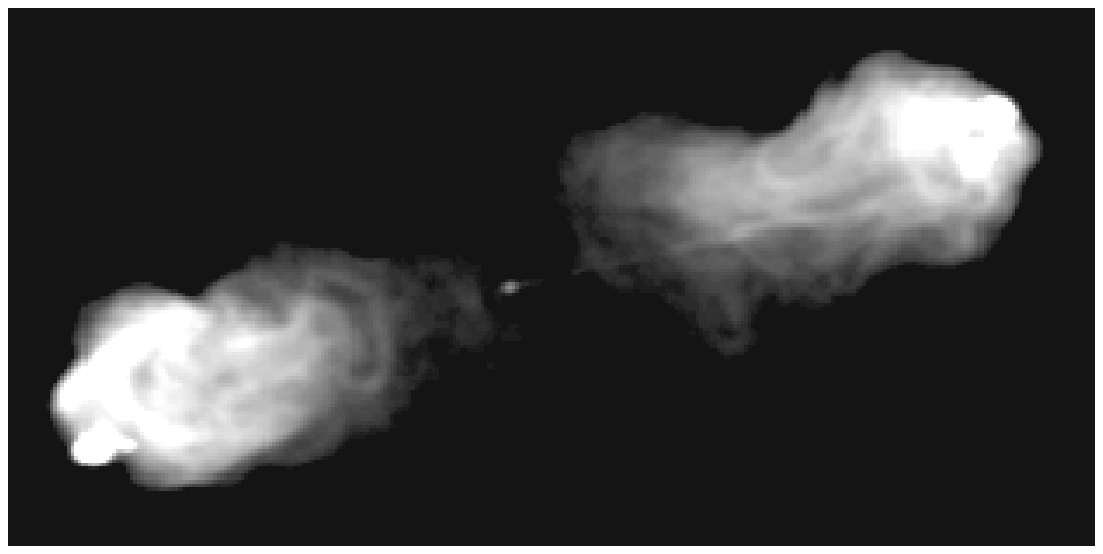,width=0.95\columnwidth}}}
\vspace{0.1in}
\dcaption{4} The classical double radio galaxy, Cygnus A.  The central
bright spot is coincident with the galactic core;  extremely well
collimated jets propagate to each side, impacting the ambient plasma
{\it via} shocks at the outer radio-bright spots.  VLA image courtesy
of C. Carilli and R. Perley (NRAO/AUI). 
\end{figure}

\paragraph{Jets with saturated instabilities}
Alternatively, if conditions (such as density contrast or Mach number) are
different, instabilities may affect the jet close to its origin.  
If the jet is subsonic, KH instabilities might lead to the development
of turbulence, causing the jet gradually to widen and entrain
local cluster plasma.  A few RG's 
do show the gradual, smooth broadening which this model predicts.  In
many others, however, the jets 
show a sudden and dramatic change, starting narrow and well collimated,
then quickly flaring out to become a broader, but again stable, flow.
Figure 5 shows one example. These ``tailed''
sources are not yet understood, but it may be that they represent
the sudden onset of an instability which saturates without fully
disrupting the flow.

\begin{figure}[htb]
\vspace{0.1in}
\centerline{ {\psfig{figure=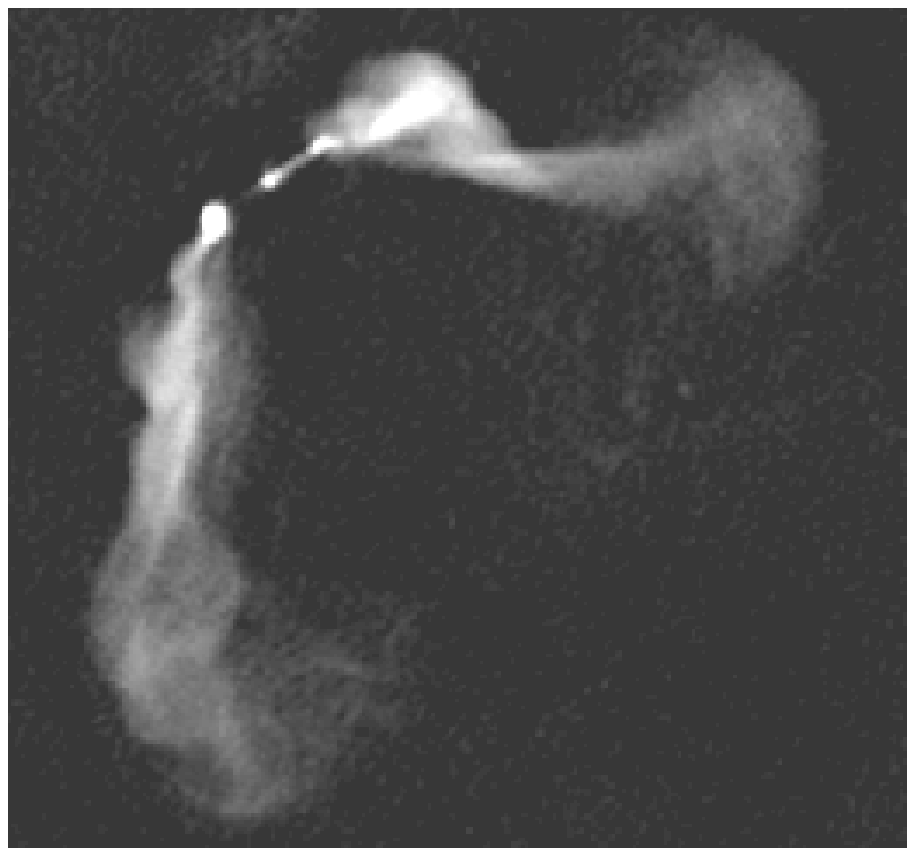,width=0.95\columnwidth}}}
\vspace{0.1in}
\dcaption{5} The tailed radio galaxy, 3C465, in a VLA radio image.$^{17}$
  Two narrow, straight
jets emanate from the galactic core (central bright spot in the
upper left).  These jets apparently go unstable, forming two
more radio-loud spots close to the core, after which the flow continues
in a more diffuse manner.  The image $\sim 200$ kpc on a side. 
\end{figure}

\paragraph{Jets which disrupt strongly} 
A third possibility is that the jet is strongly disrupted close to
its origin, and cannot stabilize again.  In this case the matter and
energy flowing in the jet will be decollimated, impacting the local
plasma in a more diffuse flow.  One expects the RG to evolve more
as a ``bubble'', growing with time as energy enters from the center,
possibly with plasma mixing across its outer edge.  A few
such RG's are known, all in special environments:  they are
attached to the central galaxy in a cooling core. We must surmise
that something unusual in conditions there leads to this disruption.
An example is the jet in M87, which disrupts within a few kpc of
the galactic core, creating the halo source shown in Figure 6. 

\begin{figure}[htb]
\vspace{0.1in}
\centerline{ {\psfig{figure=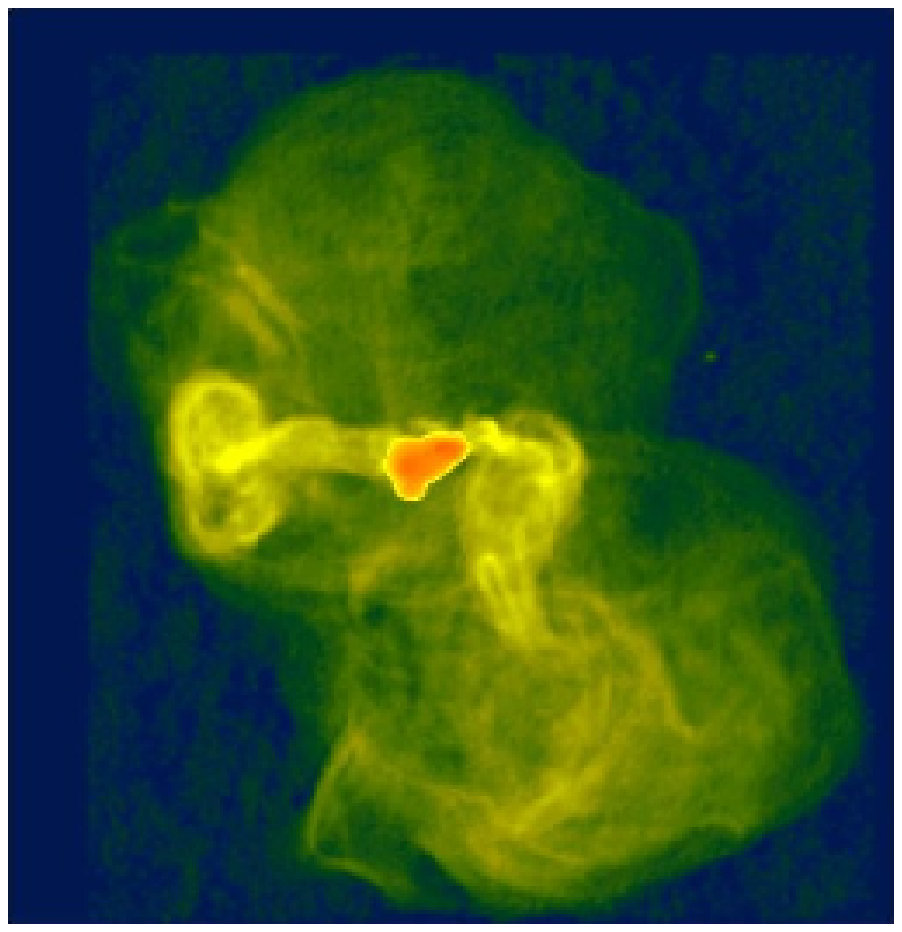,width=0.90\columnwidth}}}
\vspace{0.1in}
\dcaption{6} The large-scale radio halo in M87, in an image from the 
VLA.$^{30}$  The inner jet, shown
in Figure 3, disrupts very close to the galactic core.  The plasma
flow appears to continue in a much less ordered fashion, forming
this ``bubble'' which has partly mixed with the ambient thermal
plasma. The image $\sim 80$ kpc on a side. 
\end{figure}

\subsubsection{Impact on the ICM?}

In order to understand how the jet affects the ICM, we need to
know its power (equation 6), and also if and where that power is
deposited in the ICM.  While this last question is currently under
active debate, some general comments can be made.
Jets which disrupt strongly, and form ``bubble'' RG's, must
be interacting strongly with the local ICM.  We expect the strongest
and most localized heating in these cases. Jets which remain stable
or quasi-stable out to extragalactic scales may tend to push the ambient 
plasma aside, but not mix with it strongly (except for 
shear-surface or mixing instabilities along the lobes or tails).$^{38}$
We might expect weaker {\it local} heating of the ambient plasma in
these cases, although the jet and cluster plasmas must eventually mix.
The energy deposition may simply occur on larger scales for these
types of RG's.

\section{Issues and Controversies}

Although we have known about the radio jets and the cluster plasma
for about 30 years (since they were detected with radio interferometers
and X-ray satellites, respectively), the physical picture is still
unclear.  I present a few problems currently under discussion.

\subsection{Transport Physics}

It is not clear whether collisional transport can be applied to
all problems in the cluster environment. 

In the context of cooling flows, many authors have noted that 
heat transport at the rate allowed by Spitzer conductivity
would stop the putative strong cooling and inflow.  Many cooling
flow models simply ignored thermal conduction.  Some authors
suggested that magnetic fields could lower the conduction orders of
magnitude below the Spitzer value (as needed to preserve the early
models).  Other authors modelled heat transport assuming various
descriptions of a disordered magnetic field.  No clear consensus has
emerged but most workers find a tangled field will lead to only a
modest suppression of heat flow relative to the Spitzer value.$^{39}$. 
There has also been some work on suppression of heat flow by plasma
instabilities, again without a clear consensus.$^{40}$.

Changing views on cooling cores, in particular the inability to find
the large reservoir of cooled gas, may make this problem seem less critical.
However, the new detection of cold fronts again raises the issue.  These
fronts are thinner than the Coulomb mean free path, which is
$\sim 10$ kpc in the cluster plasma. They must be protected from
heat transport at the Spitzer level;  indeed, they must be collisionless
discontinuities.  They have been modelled in terms of strongly sheared
magnetic fields, parallel to the cold front$^{15}$;  I am not aware
of any modelling addressing whether plasma turbulence might
 support such a thin front.

\subsection{Magnetic Field Maintenance}

There are many questions yet to answer here.
We do not even know the structure of the cluster magnetic
field.  Is it strongly filamented? Is it enhanced local to embedded 
radio galaxies?  Have we correctly identified its order scale at
the small, $\sim 10$ kpc value from the Faraday observations?  Or is it
not a coincidence, that this scale is similar to the Coulomb mean free path?

We must also ask about the origin and maintenance of the field.
Some authors envision the field as primordial, enhanced by the
collapse of the cluster.  However, if
the order scale of the field is $\sim 10$ kpc, the turbulent
dissipation time will be short compared to the cluster life.
One possibility is that
turbulence or flows in the cluster support the field.  Turbulence 
 can take a small seed field (say from early-epoch galaxy ejecta) and,
given enough time, enhance it to reach approximate equipartion with
the kinetic energy of the flow.$^{24,41}$
An alternative view is that the field and flux have been
injected from active galaxies.$^{42}$  The second picture may
have trouble explaining how the magnetized jet plasma can mix effectively
with the cluster plasma to maintain fields throughout the cluster.  The
first picture is appealing to me, but our lack of understanding
of turbulence in the ICM has made it hard to test quantitatively.

\subsection{Particle acceleration}

From cosmic rays to clusters of galaxies, this is an important topic in
plasma astrophysics.  How does the plasma maintain a significant,
nonthermal population of highly relativistic particles, against radiative
losses and thermalizing collisions?  There must exist a mechanism
which preferentially energizes particles already at high energies;
otherwise we would simply see plasma heating.

There are several ``usual culprits''.  {\it Shock acceleration} can
occur {\it via} Fermi acceleration, that is
multiple scatterings, across the shock face, in 
quasi-parallel shocks;  or along the shock face,
using the induced potential drop, in quasi-perpendicular shocks.
{\it Plasma turbulent} acceleration has been proposed in many forms.
This is a stochastic process, in which fluctuating electric fields in 
plasma waves slowly energize the particles.  For relativistic particles,
the cyclotron resonance with Alfven waves is usually proposed;  but transit
time damping in strong turbulence has also been suggested.
{\it Electrodynamic} acceleration is possible if an ordered electric
 field exists, such as in current sheets or double 
layers.  {\it Reconnection} acceleration is a variant of this last.

Acceleration in radio galaxies has been an ongoing question.  Both
shock and turbulent acceleration have been suggested.
Some of the issue is determining from the dynamics what acceleration 
sites can exist away from the black hole. Shocks can be identified 
in the outer 
hot spots in classical double RG's (as in Figure 4);$^{3}$  some authors
suggest the bright knots in jets (as in Figure 3) are also shocks.$^{43}$
However, there is not always evidence for an ordered shock at acceleration
sites. Other authors$^{44}$ propose 
turbulent acceleration, possibly related to the outer shear layer of
the jet. 

The case is less clear for the cluster plasma, because it is not yet
established whether nonthermal plasma is common or rare.  Several
authors (like myself) believe it will turn out to be common, in which
case {\it in situ} acceleration is also needed in the cluster.  Once
again both shock acceleration$^{45}$ and turbulent acceleration$^{46}$
 have been invoked.  Probably both are occurring.
Large, ordered shocks are expected
in major mergers, while smaller shocks as well as turbulence are
expected more generally in smaller mergers.  Because smaller mergers
seem to be so common, and particle acceleration a direct consequence,
it seems likely that the ICM in most clusters must contain a relativistic
component.

\subsection{Energy Sources for the ICM}

Our picture of clusters has changed.  Early work in the field
assumed that clusters were static, self-contained entities in which
nothing much had happened since their creation.   We now know that
they are dynamic and still evolving.  
In particular, the cluster plasma is being energized from within
and without.  We can identify three drivers operating today;
  the challenge is to quantify them.$^{17}$

The first comes from normal galaxies within
the cluster.  As they move through the cluster plasma, they drive
turbulence and possibly bow shocks in that plasma.  This must be
operating in every cluster.  Because galaxy counting and galaxy masses
are well understood, this effect can be estimated quantitatively; 
it turns out to be only of modest importance.

A second driver is active galaxies in the cluster.  As discussed
above, the mass and energy driven out in their jets must eventually
be deposited in the ICM.  These galaxies are rare, only about one per
rich cluster, with some tendency to be located in the core.  In addition
to the jet power, we must determine the duty cycle of an AGN.  Models of
radio galaxies suggest they are short-lived compared to the cluster;
it may be that the galactic core is ``active'' only part of the time.
These problems are not yet well understood.

Finally, the ICM must be energized at present by the motions of clumps
of dark matter which continue to merge with the cluster.  General
cosmological studies, which constrain how often clumps merge and the
energy per clump, suggest plenty of energy is available to the ICM
at the present time. However, we do not  yet understand the details.
  Numerical simulations are the best tool here; but
they are only beginning to address important details such as the
small-scale motions of the dark matter and the efficiency with which
it couples to the cluster plasma.

\section*{acknowledgements} 
My understanding of this
fascinating area has benefited  by discussions
with many colleagues, over the years,
 and the published work of many authors. Due to
space limitations here I have not been able to cite many important
contributions.  Some of this work was carried out during my sabbatical
visits to the University of Oxford, the Instituto di Radioastronomia
in Bologna,  and
the Max-Planck Institut f\" ur RadioAstronomie in Bonn.  This
work was also partially supported by NSF grant AST9720263.  Finally
I thank the DPP for support to present the review talk on which this
paper is based.


\vspace{0.2in}


\small

\section*{references}

\begin{description}

\item[$^{1}$]
C. L. Sarazin, {\it X-ray Emission from Clusters of Galaxies}
(Cambridge University Press, Cambridge, 1988).

\item[$^{2}$]
In this and all
equations in this section, I suppress physical constants, to highlight
the dependence on important parameters such as $\nu, B, T, n$.

\item
[$^{3}$]
D. S. De Young, {\it The Physics of Extragalactic Radio Sources }
(University of Chicago Press, Chicago, 2002).

\item
[$^{4}$]
G. Giovannini  and L. Feretti,  in {\it Merging Processes in Galaxy
Clusters}, edited by  L. Feretti, I. M. Gioia and G. Giovannini (Kluwer,
Dordrecht, 2002) p. 197.

\item
[$^{5}$]
C. L. Carilli  and B. G. Taylor,   Annu. Rev. Astron.
Astrophys., 40, 319 (2002).

\item
[$^{6}$]
R. Fusco-Femiano, D.  dal Fiume, L. Feretti, G. Giovannini, P.  Grandi,
G.  Matt, S. Molendi, and A. Santangelo, Astrophys. J. Lett, 513, L21 (1999);
D. Gruber, and  Y. Rephaeli, Astrophys. J.,  565, 8877 (2002);
S.  Molendi, S.  De Grandi, and R.  Fusco-Femiano, Astrophys. J. Lett, 
534, L43 (2000).

\item
[$^{7}$]
S. Bowyer, T. W. Berghofer, and E. J. Korpela, Astrophys. J. 526, 592,
(1999); C. L. Sarazin  and R. Lieu, Astrophys. J. Lett., 494, L177 (1998).

\item
[$^{8}$]
L. Feretti, R. Fusco-Femiano, G. Giovannini, and F. Govoni, 
Astron. Astrophys., 373, 106 (2002).

\item
[$^{9}$]
A. C. Fabian, Annu. Rev. Astron.  Astrophys., 32, 277 (1994).

\item
[$^{10}$]
J. Binney, in {\it The Radio Galaxy Messier 87}, edited by 
H.-J. R\"oser and
K. Meisenheimer (Springer, Berlin, 1998) p. 116.

\item
[$^{11}$]
K. Roettiger, J. M.  Stone,  and J. O. Burns,  Astrophys. J., 518, 594 (1999);
S. Schindler, Astron. Astrophys., 349, 435 (1999);  
and references therein. 

\item
[$^{12}$]
U. G. Briel, J. P. Henry, R. A. Schwarz, H. B\"ohringer, H. Ebeling,
A. C. Edge, G. D. Hartnet, S. Schindler, J. Trumper and W. Voges,
 Astron. Astrophys., 246, L10 (1991); M. Sun, S. S. 
Murray, M.  Markevitch  and A. Vikhlinin,  Astrophys. J., 565, 867 (2002).

\item
[$^{13}$]
H. Rottgering, I.  Snellen, G.  Miley,  J. P. De Jong, R.J. Hanisch and
R. Perley,  Astrophys. J., 436, 654 (1994). 

\item
[$^{14}$]
T. E.  Clarke and T. A.  En\ss lin,  ``The Radio halo in Abell 2256'',
 submitted to Astrophys. J. (2002).

\item
[$^{15}$]
A. A. Vikhlinin  and M. L. Markevitch, Astronomy Letters, 26,
495 (2002); and references therein.

\item[$^{16}$] 
G. Giovannini, L.  Feretti, T.  Venturi, K.-T.  Kim,  and P. P. Kronberg,
 Astrophys. J., 406, 399 (1993).

\item[$^{17}$] 
J. A. Eilek and F. N. Owen, Astrophys. J., 567, 202 (2002).

\item[$^{18}$] 
T. E. Clarke, P. P.  Kronberg and H. B\"ohringer,  
 Astrophys. J. Lett, 547, L111 (2001).

\item[$^{19}$] 
G. S. Hennessy, F. N.  Owen, and J. A. Eilek,  Astrophys. J., 347, 144 (1989).

\item[$^{20}$] 
H. Liang, R. W.  Hunstead, M.  Birkinshaw and P. Andreani,  Astrophys. J.,
544, 686 (2002). 

\item [$^{21}$] 
G. Giovannini  and L. Feretti,   New Astronomy, 5, 535 (2002);  in addition
a systematic VLA search of a complete sample is being carried out by 
F. Owen, T. Markovic, and myself.

\item [$^{22}$] 
U. G. Briel and J. P. Henry, in {\it A New Vision of an Old Cluster:
Untangling Coma Berenices}, edited by A. Mazure, F. Casoli, F. Durret and
D. Gerbal (World Scientific, Singapore, 1998) p. 170; 
J. O. Burns, K. Roettiger, 
M. Ledlow  and A. Klypin, Astrophys. J., 427, 87 (1994).

\item [$^{23}$] 
K.-T. Kim, P. P.  Kronberg, P. E.  Dewdney and T. L. Landecker, 
 Astrophys. J., 355, 29 (1990).

\item [$^{24}$]
E. T. Vishniac, Astrophys. J., 451, 816 (1995); J. A. Eilek, in
 {\it Diffuse Thermal and Relativistic
Plasma in Galaxy Clusters}, ed. H. B\"ohringer, L. Feretti, and P. Sch\"ucker
(Max-Planck Institut, Garching,  1999), p. 71,
 and references therein.

\item [$^{25}$]
R. V. E. Lovelace, H.  Li, A. V. Koldoba, G. V.  Ustyugova,  and
M. M. Romanova,  Astrophys. J., 572, 445 (2002); D. L. Meier, S.  Koide and
U. Uchida, Science, 291, 84 (2001). 

\item[$^{26}$]
Jet deceleration will partially offset these losses, but this can be shown
quantitatively not to solve the entire problem. 

\item [$^{27}$]
F. N. Owen, P. E.  Hardee,  and T. J. Cornwell, Astrophys. J., 340, 698
(1989).

\item [$^{28}$]
H. L. Marshall, B. P.  Miller, D. S.  Davis, E. S.  Perlman, M. Wise,
C. R.  Canizares  and D. E. Harris,  Astrophys. J., 564, 683 (2002); 
E. S.  Perlman, J. A.  Biretta, W. B.  Sparks, F. D.  Macchetto,  and 
J. P. Leahy, Astrophys. J.  551, 206 (2001).

\item [$^{29}$]
J. A. Biretta, F.  Zhou and F. N. Owen, Astrophys. J., 447, 582 (1995).

\item[$^{30}$]
F. N. Owen, J. A.  Eilek and  N. E. Kassim,  Astrophys. J., 543, 611 (2000).

\item[$^{31}$]
A. Lobanov, P. E. Hardee  and J. A. Eilek, ``Internal structure and
dynamics of the kilo-parsec scale jet in M87'', to appear in New 
Astronomy Reviews (2003).

\item[$^{32}$]
A. J. Young, A. S. Wilson and C. G. Mundell, Astrophys. J., 579, 560 (2002).

\item[$^{32}$]
P. E. Hardee and A. Rosen,  Astrophys. J. 524, 650 (1998); P. E. Hardee,
Astrophys. J., 533, 176 (2000); and references therein.

\item[$^{34}$]
{\it e.g.},  J. E. Borovsky, Astrophys. J., 306, 451 (1986); L. C. Jafelice
and R. Opher, Mon. Not. R. Astron. Soc., 257, 135 (1992); S. Appl and M.
Camezind, Astron. Astrophys.,  256, 354 (1992).

\item[$^{35}$]
S. Appl, T.  Lery  and H. Baty,  Astron. Astrophys., 355, 818 (2002); 
T. Lery, H. Baty and S. Appl,  Astron. Astrophys.,  355, 1201 (2001). 

\item[$^{36}$]
A. Konigl and A. R. Choudhuri,  Astrophys. J., 289, 173 (1985). 

\item[$^{37}$]
J. A. Eilek, P. E.  Hardee, T.  Markovic, M. Ledlow and F. N. Owen, 
New Astronomy Reviews, 46, 327 (2002).

\item[$^{38}$]
New X-ray images of RG's in clusters are consistent with this:  {\it e.g.},
D. A. Clarke, D. E. Harris and C. L. Carilli,  Mon. Not. R. Astron.
Soc.,, 284, 981 (1997);  A. C. Fabian, J. S.  Sanders, W.  Ettori, G. B.
 Tayor, S. W.  Allen, C. S. Crawford, K.  Iwasawa  and R. M. Johnstone, 
Mon. Not. R. Astron. Soc., 321, 33 (2001).

\item[$^{39}$]
{\it e.g.}, B. D. G. Chandran, S. C.  Cowley, M.  Ivanushkina and R.
Sydora,  Astrophys. J., 525, 638 (1999);  R. Narayan  and M. V. Medvedev,
 Astrophys. J., 562, 129 (2001); and references therein.

\item[$^{40}$]
L. C. Jafelice, 1992, Astron. J., 104, 1279 (1992); S. Pistinner, A.
 Levinson and D. Eichler, 1996, Astrophys. J., 467, 162 (1996). 

\item[$^{41}$]
I. Goldman and Y. Rephaeli, Astrophys. J., 380, 344 (1991);
K. Dolag, M.  Bartelmann and H. Lesch, Astron. Astrophys., 387, 383 (2002).

\item[$^{42}$]
S. A. Colgate, H.  Li and V. Pariev,
Phys. Plasmas 8, 2425 (2001). 

\item[$^{43}$]
G. V. Bicknell and M.C. Begelman, Astrophys. J., 467, 597 (1996).

\item[$^{44}$]
K. Meisenheimer, ``Particle Acceleration in Radio Jets'', 
to appear in New Astronomy Reviews (2003).

\item[$^{45}$]
T. A. En\ss lin, P. L.  Biermann, U.  Klein and S. Kohle, Astron. Astrophys.,
332, 395 (1998);   F. Miniati, T. W.  Jones, H.  Kang and
D. Ryu,Astrophys. J., 562, 233 (2001).

\item[$^{46}$]
G. Brunetti, G. Setti, L.  Feretti and G. Giovannini,  Mon. Not. R.
Astron. Soc.,  320, 365 (2001); J. A. Eilek  and J. C. Weatherall,
in {\it Diffuse Thermal and Relativistic
Plasma in Galaxy Clusters}, edited by H. B\"ohringer, L. Feretti, and P. 
Sch\"ucker (Max-Planck Institut, Garching, 1999), p. 249; H. Ohno, M.
 Takizawa and S. Shibata, Astrophys. J., 577, 658 (2002); V. Petrosian,
 Astrophys. J., 557, 560 (2002).

\end{description}

\end{document}